\documentclass[11pt]{article}
\usepackage{titlesec}
\usepackage{amsmath,esdiff}
\usepackage{amssymb}
\usepackage{authblk}

\usepackage[normalem]{ulem}
\usepackage{bbold}
\usepackage{BOONDOX-cal}
\usepackage[colorlinks=true,linkcolor=blue,citecolor=teal,urlcolor=teal]{hyperref}%
\usepackage{geometry}

\usepackage{mathtools}
\usepackage{subcaption} 
\usepackage{url}
\usepackage{diagbox} 

\usepackage{setspace}
\usepackage{cite}
\usepackage{flushend}
\usepackage{mathrsfs}
\usepackage{hyperref}
\usepackage{graphicx}
\usepackage{cancel}
\usepackage{balance}
\usepackage{array,esint}
\usepackage[most]{tcolorbox}
\graphicspath{}
\titleformat{\section}{\fontsize{12}{12}\bfseries}{\thesection}{1em}{}
\twocolumn
\begin{document}
\twocolumn[\begin{@twocolumnfalse}
\title{\textbf{Simpson-Visser-AdS Black Holes: \\ Thermodynamics and Binary Merger}}
\author{\textbf{Neeraj Kumar${}^{a ~*}$, Ankur Srivastav${}^{b~\ddagger}$, Phongpichit Channuie${}^{a~ \dagger}$}}
\affil{{${}^{a}$ School of Science, Walailak University}\\
{Nakhon Si Thammarat, 80160, Thailand}\\
{${}^{b}$ Vahrenwalder Str., Hannover-30165, Germany}\\
{}}
\date{}
\maketitle
\begin{abstract}
\noindent In this article, we performed Simpson-Visser (SV)-regularization scheme to Anti-de Sitter (AdS) black holes and then studied thermal properties of the resulting spacetime geometry. We considered the validity of the first law of black hole thermodynamics in this case and derived an entropy formula consistent with this new regular geometry. Next, we carried out the free energy analysis and studied the phase structure of these black holes. We discovered non-trivial phase transition properties dependent on the SV-regularization parameter. We  also considered the validity of the second law of black hole thermodynamics and analyzed a merger scenario of two equal mass SV-regular black holes. In particular, we investigated the impact of the SV-regularization parameter on the constraints on post-merger black hole mass. Interestingly, we found that the bounds initially increase and then fall sharply with increasing the SV-regularization parameter. All results are compared with standard black holes for vanishing SV-regularization parameter.  
\end{abstract}
\end{@twocolumnfalse}]
\section{Introduction}
\noindent\let\thefootnote\relax\footnote{{}\\
{$*$nkneeraj06@gmail.com}\\
{$\ddagger$ankursrivastavphd@gmail.com}\\
{$\dagger$}phongpichit.ch@mail.wu.ac.th}

%\RepAS{}{(3) Regular BH Same thing. (4) Check \mbox{$S_n$} inequalities for \mbox{$\Lambda=0$}. (1 Done) Extension to AdS Regularization (2 Done) Black hole merger with regular entropy and variation with parameter a.} 

\noindent Black holes in general relativity are singular solutions to Einstein equations. According to Penrose's theorem \cite{Penrose, Penrose1}, a singularity is characterized by geodesic incompleteness, and it is inevitable if suitable energy conditions are respected. However, the singular nature of these objects is conceptually inconvenient, as any theory plagued with such infinities lose predictability. Since the quantum effects become prominent at high curvature scale, a complete theory that consistently integrates quantum effects into classical geometry should also remove the singular nature of the spacetime. It should be noted though that there do exist several ways to consider such a complete theory, each having its own advantages as well as shortcomings, see e.g. \cite{Carlo,kiefer01,kiefer02}.     \

\noindent However, even without referring to such a complete theory, various proposals have been put forward to circumvent the singularity issue of the spacetime while still remaining in the classical regime . The very first step in this direction was taken by Bardeen in \cite{Bardeen}, where the asymptotic mass of the black hole is replaced with a \(r\)-dependent factor in order to make the Kretschmann scalar finite. Such models still have a coordinate singularity, which represents the surface of no return for a signal propagating towards the centre, and thus the notion of an event horizon still exists. These models, without a central singularity and an event horizon, are termed as \textit{Regular} black holes in literature. Since Bardeen's proposal, various similar models of regular black holes have been proposed and studied in details \cite{reg1, reg2, reg3, reg4, reg5}. For recent review see \cite{ reg6}.  \

\noindent In 2019, Simpson and Visser \cite{SV} proposed a minimalistic regularization scheme to deal with the central singularity of the standard Schwarzschild geometry. The resulting spacetime geometry is shown to represent a regular black hole or a traversable wormhole, depending on the choice of the free parameter inserted to regularize the central singularity. It has also been shown there that the non-zero components of the Riemann tensor and curvature invariants, such as Kretschmann scalar and Weyl scalar, are finite over the whole coordinate domain. As expected from the cases of other regular black hole solutions, the associated stress-energy tensor violates the null energy condition in this case as well. \

\noindent So far, the SV-regularization scheme has not been extended for the case of AdS black holes in the literature. The motivation to study this in AdS spacetime comes from the fact that the regularization parameter is expected to modify the thermal properties of the horizon. Thus, the thermal behaviour, especially, the phase structure of the AdS black hole will modify significantly. \

\noindent In present work, we implement the SV-regularization scheme to AdS black hole spacetime. Next,
we analyse the thermal properties of the black hole and study its phase structure. In standard scenario, we find the Hawking-Page phase transition from thermal AdS to AdS black hole phase, as the Hawking temperature is tuned. Hence, it is interesting to study the phase structure of the SV-regular black hole in AdS spacetime. In this endeavour, we assume that the standard laws of black hole thermodynamics are applicable for SV-regular black holes as well.\ 

\noindent Usually, one would expect modifications to the entropy formula for regularized spacetime geometries, and it would be interesting to study the implication of the new entropy formula. In this context, we discuss a binary black hole merger scenario assuming the validity of the second law of black hole thermodynamics for SV-regular black holes, and study the constraints on the final black hole mass in both AdS spacetime and its asymptotically flat limit (that is, \(l\rightarrow\infty\)). \

\noindent The article is organized in the following way. In section (2), we review the SV-regularization scheme and applied it to write a metric for the SV-regularized AdS (SV-AdS) spacetime. Thermodynamic properties of these black holes are studied in section (3). Next, in section (4), we consider black hole merger scenario of two equal sized SV-AdS regular black holes in order to study the bounds on the mass of the resulting black hole geometry. Results are discussed in section (5). 

\section{SV-AdS Regular Black Holes}

\noindent  Simpson and Visser devised a simple regularization procedure \cite{SV} with one parameter such that it lifted the spacetime singularity and rendered any black hole solution with a regular geometry everywhere. According to this procedure, in the Schwarzschild coordinates (\(t, r, \theta, \phi \)), the coordinate \(r\) is replaced by \(\sqrt{r^2+a^2}\). Here, \(a\) is some real parameter which we shall refer to as SV-regularization paramater in the present article. This leads to extending the domain of \(r\) from \(r\in (0,+\infty)\) to \(r\in (-\infty,+\infty)\) for some non-zero value of \(a\). A Schwarzschild black hole metric with this regularization is given by 
\begin{equation}
    ds^2=-f(r,a)dt^2+\frac{dr^2}{f(r,a)}+(r^2+a^2)d\Omega^2,
    \label{geometry}
\end{equation}
where
\begin{equation}
    f(r,a)=1-\frac{2M}{\sqrt{r^2+a^2}}~.
\end{equation}
Here, \(M\) is the ADM mass and \(d\Omega^2\) is a metric on unit 2-sphere. The black hole horizon radius is given by \(f(r_h, a)=0\), that is, \(r_h=\pm\sqrt{(2M)^2-a^2}\). It is clear that the metric represents a regular black hole for \(a< 2M\) \cite{SV}. For \(a\geq2M\), the geometry becomes a wormhole. \

\noindent We shall now extend it further to include the AdS term in the lapse function. The regularization procedure remains the same. The lapse function with AdS term will become
\begin{equation}
    f(r,a)=1-\frac{2M}{\sqrt{r^2+a^2}}+\frac{r^2+a^2}{l^2}~,
    \label{lap1}
\end{equation}
where \(l\) is the AdS radius. It is related to the cosmological constant as \(\Lambda=-\frac{3}{l^2}\) for four spacetime dimensions.\

\noindent Horizon radius for the black hole is 
\begin{equation}
    r_h=\pm \sqrt{\frac{(l^2-u)^2}{3u}-a^2}~,
\end{equation}
where \(u=\left(l^6+54l^4M^2+6\sqrt{3l^8M^2(l^2+27M^2)}\right)^{1/3}\). Thus, the solution represents a regular black hole for 
\begin{equation}
\frac{(l^2-u)^2}{3u}>a^2~.
\end{equation}

\noindent In the limit \(l\rightarrow\infty\), one recovers the above mentioned SV-regularized Schwarzschild black hole condition.

\begin{table}[h!]
\centering
\setlength{\tabcolsep}{5pt}
\renewcommand{\arraystretch}{1.3}
\caption{Allowed maximum values of \(a\) for a set of \(M\) and \(l\) values for a black hole solution.}
\begin{tabular}{|c|>{\centering\arraybackslash}p{1.2cm}|>{\centering\arraybackslash}p{1.2cm}|}
\hline
\diagbox{\(M\)}{\(l\)} & 1 & \(\infty\) \\ \hline
1 & 1 & 2 \\ \hline
2 & 1.38 & 4 \\ \hline
3 & 1.63 & 6 \\ \hline
\end{tabular}
\label{t1}
\end{table}

\noindent Next, one would be interested in classical analysis of this geometry to confirm  that  it represents a regular black hole. We have carried out that analysis, using analytical expressions given for a generic static black hole solutions in \cite{curvin}, in the supplementary file attached with this article. We found that the geometry is regular everywhere and in limit, \(l\rightarrow\infty\), we recovered the expressions of curvature invariants obtained in \cite{SV}. Thus, the geometry represented by eq.{(\ref{geometry})} is a regular black hole solution.  \

\noindent In this article, we are more interested in thermodynamic properties of these geometries. This is discussed in the next section. 

\section{Thermodynamic Properties and Phase Structure}
\noindent Existence of a horizon in the spacetime geometry allows one to define the surface gravity, which is then linked to the Hawking temperature. This allows one to study the thermal properties of regular black holes. We shall start with the expression of the lapse function in eq.(\ref{lap1}) and calculate the Hawking temperature as 
\begin{align}
    T&=\frac{f'(r,a)|_{r=r_h}}{4\pi}\nonumber\\
    &=\frac{1}{4\pi}\left[\frac{r_h}{r_h^2+a^2}+\frac{3r_h}{l^2}\right]~.
\end{align}
We can express the black hole mass in terms of the horizon radius, using \(f(r_h,a)=0\), which takes the following form
\begin{equation}
    M=\frac{\sqrt{r_h^2+a^2}}{2}\left(1+\frac{r_h^2+a^2}{l^2}\right)~.
    \label{mass}
\end{equation}
Next, we assume the validity of the first law of the black hole thermodynamics and calculate the entropy associated with the SV-AdS regular black hole spacetime. Although, such an assumption is non-trivial and there are still efforts being made in order to show the compatibility of the first law with the regular black hole geometries \cite{murk}. The first law for the black holes considered here can be given as 
\begin{equation}
    dM=TdS~.
\end{equation}
Using this, one can calculate the entropy formula consistent with the first law. It takes the form 
\begin{equation}
    S=\int_0^{r_h}\frac{dM(r_h')}{dr_h'}\frac{1}{T}dr_h'~. 
\end{equation}
Solving the above integral leads us to
\begin{equation}
  \hspace{-5mm}  S=\pi\left(r_h\sqrt{r_h^2+a^2}-a^2\log{\frac{\sqrt{r_h^2+a^2}-r_h}{a}}\right)~.
\end{equation}
In limit \(a\rightarrow0\), we recover the standard Bekenstein-Hawking entropy formula. 

\begin{figure}
    \includegraphics[width=1\linewidth]{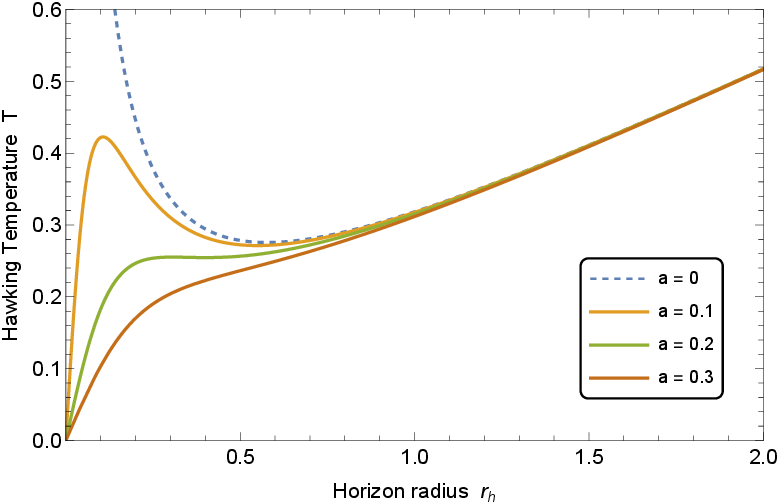}
        \caption{Hawking temperature vs horizon radius for SV-regular AdS black holes for \(a=(0, 0.1, 0.2, 0.3)\) as we set \(l=1\). }
\label{f1}
    \end{figure}

\noindent From the expression of the mass of the black hole in eq.(\ref{mass}), it is clear that it is a monotonic function of the horizon radius, thus, it increases as the black hole size increases. This behaviour is similar to the standard Schwarzschild AdS black holes. Next, we plot the Hawking temperature with the horizon radius in Fig.{(\ref{f1})}.
It is clear that the behaviour matches with the standard AdS black holes for large black holes, however, it noticeably differs for small black holes. Unlike the standard Schwarzschild AdS black holes, there is a local maxima and a minima and there is an extremal black hole limit for the SV-AdS regular black holes. Since there is a black hole for all values of temperature, there is no stable thermal AdS phase, hence, there is no Hawking-Page phase transition\cite{HawkingPage}.  \   

\noindent From the plot given in Fig.{(\ref{f1})}, it is also clear that there are two stable branches, small and large black holes, and there is an unstable intermediate branch. Interestingly, as one tunes the SV-regularization parameter, the temperature behaviour changes and for some value of the parameter there is a point of inflexion. Above this point, there exist only one stable phase for such regular black holes. Next, we carry out the free energy analysis to understand the phase structure of the SV-AdS regular black holes.   \\

\noindent \textbf{Free Energy Analysis:}\\

\noindent Free energy analysis provides all the necessary information of the phase transition properties of a thermal system. For the SV-AdS regular black holes, one can write the free energy from the above calculated thermal parameters. The expression takes the following form
\begin{align}
F &= M - TS \nonumber\\
  &= \frac{\sqrt{r_h^2+a^2}}{2}\left(1+\frac{r_h^2+a^2}{l^2}\right)
     - \frac{1}{4}\left[\frac{r_h}{r_h^2+a^2}+\frac{3r_h}{l^2}\right]
     \nonumber\\
  & \hspace{-5mm} \quad\times\left(r_h\sqrt{r_h^2+a^2}-a^2\log{\frac{\sqrt{r_h^2+a^2}-r_h}{a}}\right)~.
  \label{freeEn}
\end{align}
In limit $a\rightarrow0$, the expression for the free energy reduces to 
\begin{equation}
    F_{a\rightarrow0}=\frac{r_h}{4}\left(1-\frac{r_h^2}{l^2}\right)~.
\end{equation}
This expression matches with the free energy of a Schwarzschild AdS black hole \cite{HawkingPage}. \

\begin{figure}
    \includegraphics[width=1\linewidth]{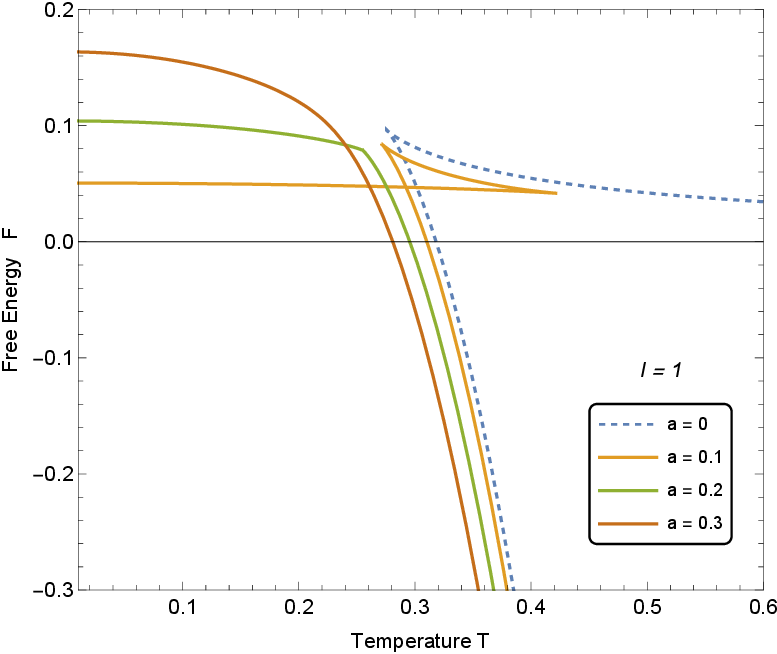}
        \caption{Free energy vs Hawking temperature for SV-AdS regular black holes for \(a=(0, 0.1, 0.2, 0.3)\) as we set \(l=1\). }
\label{f2}
    \end{figure}

\noindent We plot the free energy for the SV-AdS regular black holes, given by eq.(\ref{freeEn}), against Hawking temperature, in Fig.(\ref{f2}), for different values of the SV-regularization parameter, \(a\).
Dashed plot represents the standard Schwarzschild AdS black hole with two branches, one with positive free energy with respect to the thermal AdS (free energy negligible in comparison to the AdS black hole phase) and another one which extends to negative direction below the temperature axis. It can be seen that for large temperature, the AdS black hole state is more preferred than the thermal AdS state \cite{HawkingPage}.\

\noindent It is also evident in Fig.(\ref{f2}) that, for non-zero values of \(a\), there is no thermal AdS state as there is always a black hole solution for all temperatures. This observation is also consistent with the temperature plot given in Fig.(\ref{f1}). Here, we consider the interpretation similar to a fixed charge ensemble made in \cite{Chamblin1, Chamblin2}. In this case, the background is the extremal black hole corresponding to \(T\rightarrow0\) limit and the free energy associated with this state is finite as depicted in Fig.(\ref{f2}). We are still in canonical ensemble and free energy change is considered with respect to this assumed background. The free energy of the black hole state is negative with respect to its extremal limit and hence, it is more stable and preferred state of the system. From the plot, it is also clear that there is a first order small-to-large black hole transition as one tunes the temperature. \

\noindent As we increase the values of the parameter \(a\), the phase transition points shifts towards a critical point. This behaviour is similar to van der Waal's fluid like behaviour, and it has been observed in black hole cases in several studies \cite{RM1,RM2,SG,NK1,NK2}. Thus, it is an important observation that even a singularity regularization parameter changes the phase structure of black hole significantly. It would be interesting to study the behaviour of critical exponents near the critical point and determine the universality class these SV-AdS regular black holes belong to. Next, we shall study the black hole merger scenarios of these SV-regularized black holes. 

\section{Constraints on Black Hole Merger}
The most important modification to the black hole thermodynamics for these SV-AdS regular black holes comes from the expression of the entropy formula. Implications from such a modification has not yet been addressed in details in the literature. Through the present article, we set out to address this important issue and show that implications of such a modification to the black hole geometry are indeed profound and non-trivial. Particularly, we have explored the implications of entropy and the associated second law for the case of two equal mass SV-AdS regular black hole merger event. Bounds on the mass of the final product due to a black hole merger event in the presence of modified entropies have been discussed before \cite{Alice01, NK3}. Such studies can help in constraining the theory space.\ 

\noindent We have carried out merger event analysis for SV-regularized black holes in the current section and studied the impact of the regularization parameter, $a$, on the constraints on the final mass post-merger. We have explicitly discussed the merger scenario in asymptotically flat, that is, \(l\rightarrow\infty\) limit, and in AdS spacetime.   \\ 

\noindent \textbf{Merger Constraints on SV-AdS Regular Black Holes:} \\

\begin{figure}[h!]
    \includegraphics[width=1\linewidth]{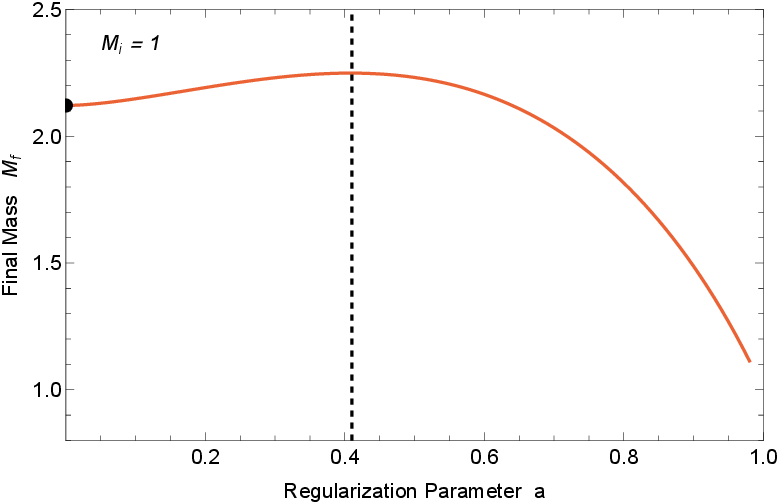}
        \caption{Final Mass vs SV-Regularization Parameter \(a\) as we set \(l=1\) and \(M_i=1\). }
\label{f3}
    \end{figure}

\noindent In this section, we have considered a merger scenario of two equal mass SV-AdS regular black holes. Initial entropy of the complete system is given by 2$S_i$, where $S_i$ is given by the following expression 
\begin{equation}
    S_i=\pi\left(r_h^i\sqrt{{r_h^i}^2+a^2}-a^2\log{\frac{\sqrt{{r_h^i}^2+a^2}-r_h^i}{a}}\right).
\end{equation}
Here, \(r_h^i\) is the horizon radius of a SV-AdS regular black hole before merger.
Similarly, if the final black hole radius is represented by \(r_h^f\) and entropy by \(S_f\) then according to the second law of black hole thermodynamics we have 
\begin{equation}
    S_f\geq 2S_i~.
    \label{secondlaw}
\end{equation}
Eq.(\ref{secondlaw}) provides a relation between the horizon radius of black holes before merger and that for the final product after merger. With this relation, we shall write the expression for the mass of the final black hole for a set of fixed values of the initial mass and plot it with the SV-regularization parameter \(a\). In Fig.(\ref{f3}), we have set initial mass, $M_i =1$ without any loss of generality.
The plot here shows that the final mass post merger initially increases as we increase the value of \(a\) and then starts decreasing after a certain value, indicated by a dashed line. Thus, the bounds on the final mass change non-trivially with respect to SV-regularization parameter, \(a\). The bound  corresponding to the standard Schwarzschild AdS black hole is denoted by a black dot on the vertical axis. \

\begin{figure}
    \includegraphics[width=1\linewidth]{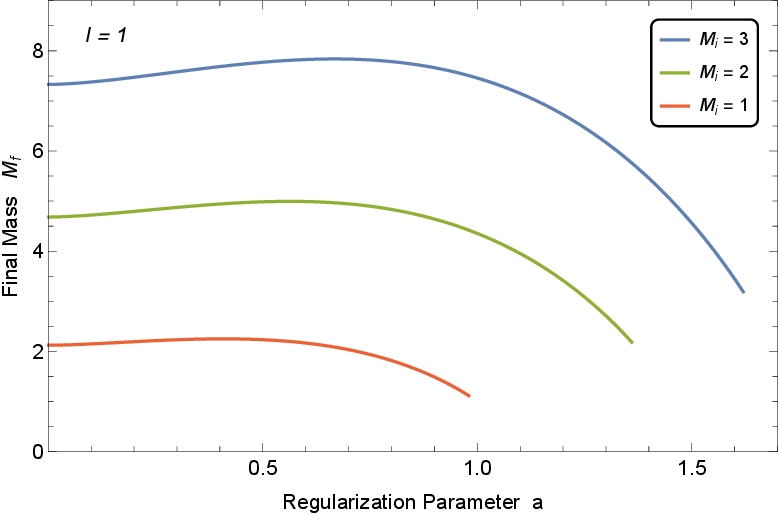}
        \caption{Final Mass vs SV-Regularization Parameter \(a\) as we set \(l=1\) and \(M_i=1,2,3\). }
\label{f4}
    \end{figure}

\noindent Next, we studied similar variations as we change the initial mass in Fig.(\ref{f4}).
The plot reveals that the qualitative feature of the variation remains same as in Fig.(\ref{f3}) while the point of maxima shifts towards right. It should be noted that the range of the SV-regularization parameter for the final product to remain a SV-AdS regular black hole solution increases too. This range, considered in the analysis, corresponds to the initial mass value \(M_i\) and it remains valid in the post merger scenario as well (that is, the solution is still a SV-AdS regular black hole post merger). Now, we shall analyse the case of SV- regularized black hole merger in asymptotically flat spacetime. \\

\noindent\textbf{Merger Constraints on SV-flat Black Holes:} \\

\begin{figure}
    \includegraphics[width=1\linewidth]{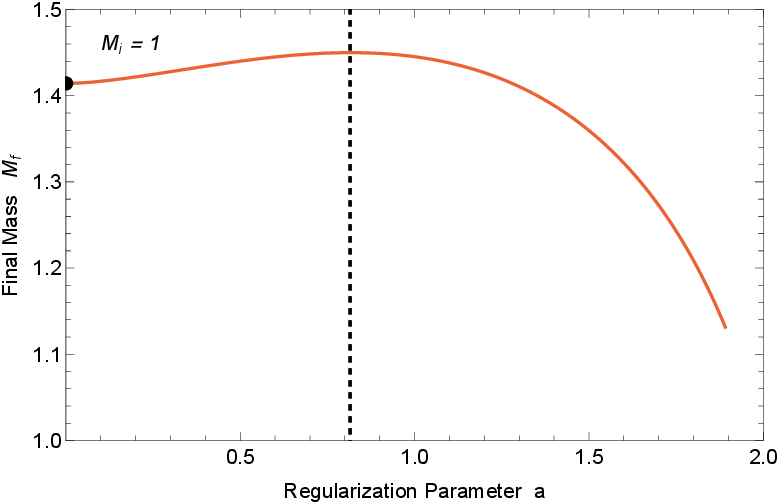}
        \caption{Final Mass vs SV-Regularization Parameter \(a\) as we set \(l=\infty\), \(M_i=1\). }
\label{f5}
    \end{figure}

\begin{figure}
    \includegraphics[width=1\linewidth]{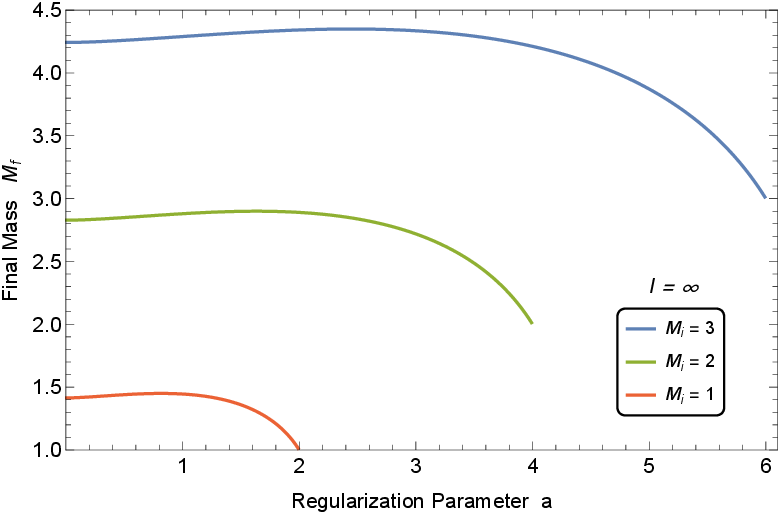}
        \caption{Final Mass vs SV-Regularization Parameter \(a\) as we set \(l=\infty\) and \(M_i=1,2,3\). }
\label{f6}
    \end{figure}

\noindent The entropy formula is independent of the AdS parameter \(l\). Thus, it remains same for the asymptotically flat spacetime. The mass of the SV-regularized black hole in asymptotically flat spacetime will be given by eq.(\ref{mass}) in limit \(l\rightarrow\infty\). The expression for the mass of the black hole then takes the form
\begin{equation}
    M=\frac{\sqrt{r_h^2+a^2}}{2}~.
\end{equation}
Now, using this mass relation and the second law in eq.(\ref{secondlaw}), we have plotted final mass post merger with the SV-regularization parameter, \(a\), in Fig.(\ref{f5}).
 The plot shows similar behaviour as for the case of SV-AdS regular black holes.  There is a value of the SV-regularization parameter, \(a\), for which the bound on the final black hole mass is maximum and it falls as \(a\) is tuned further. Next, we wish to know the behaviour as we vary the initial mass of the black holes. This is depicted in Fig.(\ref{f6}).
The behaviour remains same as we tune the initial mass of merging black holes. The range of SV-regularization parameter, \(a\), in this case increases with an increase in the initial mass. The explicit allowed values for the cases discussed in this paper are given in Table {\ref{t1}}.
    
\section{Discussions}

\noindent Singularities in general relativity are inevitable if energy conditions of ordinary matter are respected. However, these singularities also predict the demise of general relativity as a complete theory of gravity. SV-regularization scheme is a bottom-up approach which provides a way to circumvent central black hole singularity. The geometry obtained as such is free from singularities and its impacts are visible in classical and thermal regime. \

\noindent We employed the SV-regularization procedure to black holes in AdS spacetime and studied their classical properties. Classically, these geometries are free from central singularities and are regular everywhere for non-zero values of the SV-regularization parameter, $a$. We have confirmed this by explicit calculations of the curvature invariants viz. the Weyl and the Kretschmann scalars. These calculations are provided in the supplementary file attached to this article.\

\noindent Next, we studied thermal properties of these SV-regular black holes geometries. The SV-regularization parameter, \(a\), denotes divergence from the standard black hole geometries. After a confirmation that the SV-regularization procedure renders the spacetime regular, we obtained the thermal properties of these regular black holes. We assumed that the geometry follows the first law of black hole thermodynamics and obtained the expression of the black hole entropy. The entropy formula has a non-trivial dependence on the SV-regularization parameter, \(a\), and we obtained standard Bekenstein-Hawking formula in \(a\rightarrow0\) limit. The SV-regularization parameter, \(a\), also appeared in other thermodynamic expressions. One important point of divergence from the standard AdS black holes emerged in the expression of the Hawking temperature. For these SV-regularized black hole geometries, there exists an extremal black hole limit and the black hole solutions exist for all temperature range. \ 

\noindent Next important properties of black holes in the AdS spacetime are linked to their phase structure. There is a famous Hawking-Page phase transition for standard AdS black holes between the thermal AdS and the black hole phase. This structure completely modifies as the regularization parameter is turned on. Now, instead of the thermal AdS phase there is a black hole phase for all temperature range. The free energy analysis revealed that there is a small-to-large black hole phase transition as the Hawking temperature is tuned. Also, the behaviour is similar to van der Waal's fluid as there exists a critical point. \

\noindent The second law of black hole thermodynamics provides constraints on the final black hole parameters in a binary black hole merger scenario. We considered the validity of this law for SV-regular black holes and implemented it for two equal mass black holes. The non-trivial entropy expression provides interesting bounds on the final masses of the SV-regular black holes resulting from the merger. We studied merger of SV-regular black holes in both AdS and asymptotically flat spacetime. Interestingly, we discovered that the final mass bounds increase to a maximum value and then decrease as we tune the SV-regularization parameter. Thus, the maximum energy radiated in the form of gravitational wave in a merger scenario has a non-trivial dependence on the the SV-regularization parameter. This behaviour is very informative and may be utilized to put bounds on the SV-regularization parameter from the gravitational wave data. \ 

\noindent We encountered an interesting phase transition behaviour of these black holes and their similarity to van der Waal's fluid. These black holes showed critical behaviour for a certain value of the SV-regularization parameter. One would be interested in analyzing the second order phase transition at the critical point and calculate respective critical exponents associated with it in order to find the universality class of these geometries, which we have left for future works.

\section*{Acknowledgement} This research has received funding support from the NSRF via the Program Management Unit for Human Resource and Institutional Development, Research and Innovation grant number B13F680075. AS would like to acknowledge Mrs. Megha Dave for financial support during the work.

\balance


\begin{thebibliography}{8}

%\raggedright
\bibitem{Penrose}
R. Penrose, %\href{https://doi.org/10.1103/PhysRevLett.14.57 }
{Phys. Rev. Lett. 14, 57, (1965)}
\bibitem{Penrose1}
R. Penrose, %\href{https://doi.org/10.1023/A:1016578408204}
{General Relativity and Gravitation 34, 1141–1165 (2002)}
\bibitem{Carlo} C. Rovelli, %\href{https://doi.org/10.48550/arXiv.gr-qc/9803024}
{arXiv:gr-qc/9803024v3 (1998)}

\bibitem{kiefer01} C. Kiefer, %\href{https://global.oup.com/academic/product/quantum-gravity-9780199585205?cc=de&lang=en&}
{International Series of Monographs on Physics (2025)}

\bibitem{kiefer02} C. Kiefer, %\href{https://doi.org/10.48550/arXiv.2302.13047}
{arXiv:2302.13047v1 [gr-qc]}

\bibitem{Bardeen}
J. M. Bardeen, %\href{Proceedings of International Conference GR5, 1968, Tbilisi, USSR, p. 174}
{Proceedings of International Conference GR5, 1968, Tbilisi, USSR, p. 174}

\bibitem{reg1}
E. Ayón-Beato, A. García, %\href{https://doi.org/10.1016/S0370-2693(00)01125-4}
{Physics Letters B, Volume 493, Issues 1–2, (2000)}
\bibitem{reg2}
E. Ayón-Beato, A. García, %\href{https://doi.org/10.1103/PhysRevLett.80.5056}
{Phys. Rev. Lett. 80, 5056 (1998)}
\bibitem{reg3}
A. Borde, %\href{https://doi.org/10.1103/PhysRevD.50.3692}
{Phys. Rev. D 50, 3692, (1994)}

\bibitem{reg4}
S. A. Hayward, %\href{https://doi.org/10.1103/PhysRevLett.96.031103 }
{Phys. Rev. Lett. 96, 031103, 2006}

\bibitem{reg5}
V. P. Frolov, %\href{https://doi.org/10.1103/PhysRevD.94.104056 }
{Phys. Rev. D 94, 104056, 2016}

\bibitem{reg6}
C. Lan et al., %\href{https://doi.org/10.1007/s10773-023-05454-1}
{ Int J Theor Phys 62, 202 (2023)} 
\bibitem{SV}
A. Simpson, M. Visser, %\href{https://iopscience.iop.org/article/10.1088/1475-7516/2019/02/042}
{JCAP02(2019)042}

\bibitem{curvin}
F.S.N. Lobo et al., %\href{https://doi.org/10.1103/PhysRevD.103.084052}
{Phys. Rev. D 103, 084052 (2021)}

\bibitem{murk}
S. Murk, I. Soranidis, %\href{https://doi.org/10.1103/PhysRevD.108.044002}
{Phys. Rev. D 108, 044002 (2023)}

\bibitem{HawkingPage}
S. W. Hawking, D. N. Page,%\href{https://doi.org/10.1007/BF01208266}
{Commun.Math. Phys. 87, 577–588 (1983)} 

\bibitem{Chamblin1}
A. Chamblin et al. %\href{https://doi.org/10.1103/PhysRevD.60.064018}
{Phys. Rev. D 60, 064018 – Published 24 August, 1999}

\bibitem{Chamblin2}
A. Chamblin et al., %\href{ https://doi.org/10.1103/PhysRevD.60.104026}
{Phys. Rev. D 60, 104026 – Published 25 October, 1999}

\bibitem{RM1}
D. Kubiznak, R. B. Mann,%\href{https://doi.org/10.1007/JHEP07(2012)033}
{ J. High Energ. Phys. 2012, 33 (2012)} 

\bibitem{RM2}
D. Kubiznak et al., %\href{https://iopscience.iop.org/article/10.1088/1361-6382/aa5c69}
{Class. Quantum Grav. 34 063001 (2017)}

\bibitem{SG}
S. Gunasekaran, et al., %\href{https://doi.org/10.1007/JHEP11(2012)110}
{J. High Energ. Phys. 2012, 110 (2012)}

\bibitem{NK1}
N. Kumar et al., %\href{https://doi.org/10.1103/PhysRevD.106.026005}
{Phys. Rev. D 106, 026005 (2022)}

\bibitem{NK2}
N. Kumar et al., %\href{ https://doi.org/10.1103/PhysRevD.107.046005}
{Phys. Rev. D 107, 046005 (2023)}


\bibitem{Alice01}
Alice Bernamonti et al., %\href{https://doi.org/10.1007/JHEP10(2024)177}
{J. High Energ. Phys. 2024, 177 (2024)}

\bibitem{NK3}
N. Kumar et al., %\href{https://doi.org/10.48550/arXiv.2509.08362}
{arXiv:2509.08362v1 [gr-qc] (2025)}



\end{thebibliography}
\end{document}